\documentclass[letterpaper, 10 pt, conference]{ieeeconf}  

\IEEEoverridecommandlockouts                              
\usepackage{graphicx}
\usepackage{etoolbox}
\usepackage[colorlinks]{hyperref}
\usepackage{listings}
\usepackage{amstext}
\usepackage{bbm}
\usepackage{amsmath}
\usepackage{amsfonts}
\usepackage{amssymb}
\usepackage{xspace}
\usepackage{amsmath,array,graphicx}
\usepackage{kantlipsum}
\usepackage{multirow}

\usepackage{color}

\makeatletter
\patchcmd{\@verbatim}
  {\verbatim@font}
  {\verbatim@font\scriptsize}
  {}{}
\makeatother

\title{\LARGE \bf
A Deep Learning Approach to Fast, Format-Agnostic Detection of Malicious Web Content
}

\author{Joshua Saxe$^{1}$, Richard Harang$^{2}$, Cody Wild$^{3}$, Hillary Sanders$^{4}$
\thanks{*This research was conducted at Sophos}
\thanks{$^{1}$
        {\tt\small joshua.saxe at sophos.com}}%
\thanks{$^{2}$
        {\tt\small richard.harang at sophos.com}}%
\thanks{$^{3}$
        {\tt\small cody.wild at sophos.com}}%
\thanks{$^{4}$
        {\tt\small hillary.sanders at sophos.com}}%
}

\begin{document}

\maketitle
\thispagestyle{empty}
\pagestyle{empty}

\begin{abstract}
Malicious web content is a serious problem on the Internet today. In this paper we propose a deep learning approach to detecting malevolent web pages. While past work on web content detection has relied on syntactic parsing or on emulation of HTML and Javascript to extract features, our approach operates directly on a language-agnostic stream of tokens extracted directly from static HTML files with a simple regular expression. This makes it fast enough to operate in high-frequency data contexts like firewalls and web proxies, and allows it to avoid the attack surface exposure of complex parsing and emulation code. Unlike well-known approaches such as bag-of-words models, which ignore spatial information, our neural network examines content at hierarchical spatial scales, allowing our model to capture locality and yielding superior accuracy compared to bag-of-words baselines. Our proposed architecture achieves a 97.5\% detection rate at a 0.1\% false positive rate, and classifies small-batched web pages at a rate of over 100 per second on commodity hardware. The speed and accuracy of our approach makes it appropriate for deployment to endpoints, firewalls, and web proxies.

\end{abstract}

\section{Introduction}

Malicious web content is a major element in cyber-attacks observed today.  This harmful content comes in two categories.  The first category involves attacker-crafted web content that exploits browser software vulnerabilities to achieve malicious ends on users computers.  The second category, phishing, targets human fallibility, and consists of web content that tricks users into inadvertently divulging financial information or login credentials.  Attacks leveraging malicious content are highly prevalent on today's web.  For example, in one week in December 2017, more than 100,000 previously unseen malicious HTML files were observed on a threat intelligence aggregator.  
%
%
%
%

There are multiple challenges to detecting and blocking this kind of content.  First, detection approaches must operate quickly on the commodity hardware used in user endpoints and firewalls, so that they do not slow down users' browsing experience.  Second, approaches must be resilient to syntactic and semantic changes in malicious web content, so that adversarial evasion techniques like Javascript obfuscation and text randomization do not fly under the radar.  Finally, detection approaches must be able to find “needles in the haystack”: small snippets of code embedded in otherwise benign web content, which indicate that a page is dangerous.  This is important because many of today's web attacks are delivered via ad networks or comment feeds as small components of otherwise benign webpages.
%
%
%
%

To address these challenges, this paper proposes a deep learning approach to detecting malicious web content.  Our model uses a simple, fast, 12 character regular expression to tokenize web content, and then examines this content at multiple hierarchical spatial scales. ``Hierarchical spatial scales" here means that instead of simply using token aggregation over the full document as an input, we calculate representations that aggregate over multiple locally specific subregions: dividing the document into halves, quarters, eights, and sixteenths. We then apply two dense layers - which we refer to as the inspector network - over all these levels of aggregation to extract a representation of the document at these multiple spatial scales.
%
%
%
%
%
%

We compare our proposed approach to a number of baselines, including simple bag-of-words models and more complex deep architectures, and show that it achieves the best results at a reasonable computational cost.    We achieve a detection rate of more than 97\% at a false positive rate of 0.1\% on temporally disjoint, previously unseen content, without the extra complexity of parsing of web content or emulating its behavior.  Indeed, our results suggest that malicious content detection models utilizing deep learning can learn high quality representations of web content based on a simple stream-of-tokens input.
%
%
%
%
%
%

The structure of the rest of this paper is as follows.  In section \ref{sec:prevwork} we review related work in the areas of malicious HTML, Javascript and URL detection, as well as related work in deep learning and natural language processing literature.  In section \ref{sec:method} we lay the groundwork intuitions for our hierarchical architecture choice, and describe how it works on a mechanical level.  Subsequently, in \ref{sec:results} we describe our experimental setup, detail the experiments we ran,  and provide analysis of our experimental results.
%
%

\section{Previous Work}
\label{sec:prevwork}

Our work relates to research in the areas of heuristic web content detection, machine learning web content detection, and deep learning document classification.  Below we review this work and place it in dialogue with our own.

One body of web detection work focuses solely on using URL strings to detect malicious web content.  \cite{ma2009beyond} proposes a machine learning system for detecting malicious URLs.   They focus on using manually defined features to maximize detection accuracy.  \cite{saxe2017expose} also focuses on detecting malicious web content based on URLs, but whereas the first of these uses a manual feature engineering based approach, the second shows that learning features from raw data with a deep neural network achieves better performance.  \cite{choi2011detecting} uses URLs as a detection signal, but also incorporates other information, such as URL referrers within web links, to extract hand-crafted features which they provide as input to both SVM and K-nearest neighbors classifiers.
%
%
%
%
%
%
%
%
%
%
%
%
%
%
%
%
%
%

All of these approaches share a common goal with our work, the detection of malicious web content, but because they focus only on URLs and related information, they are unable to take advantage of the malicious semantics within the web content.  While URL-based systems have the advantage of being lightweight and can be deployed in contexts where full web content is not available, our work focuses on HTML files because their richer structure and higher information content. Since these approaches use orthogonal input information, there is certainly room for HTML-based and URL-based approaches to be combined into an even more effective ensemble system. 
%
%
%
%
%
%
%
%
%
%
%
%
%

A body of work including \cite{canali2011prophiler}, \cite{seifert2008identification}, \cite{kumar2011efficient} and \cite{feinstein2007caffeine} attempts to detect malicious web content by manually extracting features from HTML and Javascript, and feeding them into either a machine learning or heuristic detection system.  \cite{canali2011prophiler} proposes an approach that extracts a wide variety of features from a page's HTML and Javascript static content, and then feeds this information to machine learning algorithms. They try several combinations of features and learning algorithms and compare their relative merits.  \cite{seifert2008identification} eschews machine learning and proposes manually-defined heuristics to detect malicious HTML, also based on its static features.  \cite{kumar2011efficient} also utilizes a heuristic-based system, but one which leverages both a Javascript emulator and HTML parser to extract high quality features.  Similarly, \cite{feinstein2007caffeine} proposes a web crawler with an embedded Javascript engine for Javascript deobfuscation and analysis to support detection of malicious content.
%
%
%
%
%
%
%
%
%
%
%

The approach we propose here is similar to these efforts in that we focus on detailed analysis of HTML files, which include HTML, CSS, and embedded Javascript. Our work differs in that instead of parsing HTML, Javascript or CSS explicitly, or emulating Javascript, we use a parser-free tokenization approach to compute a representation of HTML files.  A parser-free representation of web content allows us to make a minimal number of assumptions about the syntax and semantics of malicious and benign documents, thereby allowing our deep learning model maximum flexibility in learning an internal representation of web content.  Additionally, this approach minimizes the exposed attack surface and computational cost of complex feature extraction and emulation code.  
%
%
%
%
%
%
%
%

Outside of the web content detection literature, researchers have made wide-ranging contributions in the area of deep learning based text classification.  For example, in a notable work, \cite{dos2014deep} shows that 1-dimensional convolutional neural networks, using sequences of both unsupervised (word2vec) and fine-tuned word embeddings, give good or first-rank performance against a number of standard baselines in the context of sentence classification tasks.  \cite{zhang2015character} goes beyond this work to show that CNNs that learn representations directly from character inputs perform competitively relative to other document classification approaches on a range of text classification problems.  Relatedly, \cite{dos2014deep} proposes a model that combines word and character-level inputs to perform sentence sentiment classification.

Our work relates to these approaches in that, because our model uses a set of dense network layers that apply the same parameters over multiple subdivisions of a file's tokens, it can be interpreted as a convolutional neural network that operates over text.  One difference between our work and the work cited above is that we do not operate on natural language exclusively.  Instead, our architecture operates on HTML documents that are, in practice, a mash-up of HTML, Javascript, and CSS, with arbitrary source code, attack payloads, and natural language expressed within these formats.  Because in-the-wild HTML content makes defining a discrete vocabulary of tokens difficult, we do not use word embeddings as our model input.  Instead we use a novel hierarchical representation of web documents based on simple, format-agnostic tokenization.
%
%
%
%
%
%
%
%

Also, unlike some of the past work on deep learning based document classification, we avoid using raw character sequences as input to our model.  We do this because the length of typical HTML documents makes inference involving convolutions or recurrent architectures over raw character strings intractable on today's commodity endpoints and firewalls.  Our work also differs from most document classification work in that we seek to detect an active adversary attempting to evade detection (in contrast to, for example, sentiment classification, in which sentence authors do not seek to evade detection of the sentiments they express).
%
%
%
%

\section{Methodology}
%
%
%
\label{sec:method}

\subsection{Intuitions and design principles}
%
%
\label{sec:intuitions}

A number of intuitions, which we list below, motivate the model we propose in this paper:
\begin{enumerate}
\item Malicious web pages often have small snippets of malicious content (such as malicious Javascript) embedded in some variable amount of benign content (e.g. the original benign content in a hacked webpage). Identifying that a given document is malicious thus requires that a model examine the document at multiple spatial scales. That is because the range of sizes of malicious Javascript snippets is small, but the variance in length of HTML documents is quite large, meaning that proportion of the document length that represents malicious content is variable among examples.
%
%
%
%
%
%
\item Explicit parsing of HTML documents, which in reality are collections of HTML, Javascript, CSS, and raw data, is undesirable, because it significantly complicates implementation, can require high computational overhead, and opens up an attack surface within the detector itself which could potentially be exploited by attackers.
\item Emulation, static analysis, or symbolic execution of Javascript within HTML documents is undesirable because of the computational overhead it imposes and because of the attack surface it opens up within the detector.
\end{enumerate}

Following from these intuitions, we made the following high level design decisions in creating our proposed approach:

\begin{enumerate}

\item Instead of performing detailed parsing, static analysis, symbolic execution, or emulation of content within HTML documents, we compute a simple bag-of-words-style tokenization of documents that makes minimal assumptions about their constituent formats.
\item Instead of simply using a flat, bag-of-tokens representation aggregated over the entire document, we use a representation that captures locality at multiple spatial scales representing different levels of localization and aggregation, allowing our model to find “needle-in-the-haystack” malicious content within otherwise benign malicious documents.
\end{enumerate}

\begin{figure}
%
%
%
      \includegraphics[width=\linewidth]{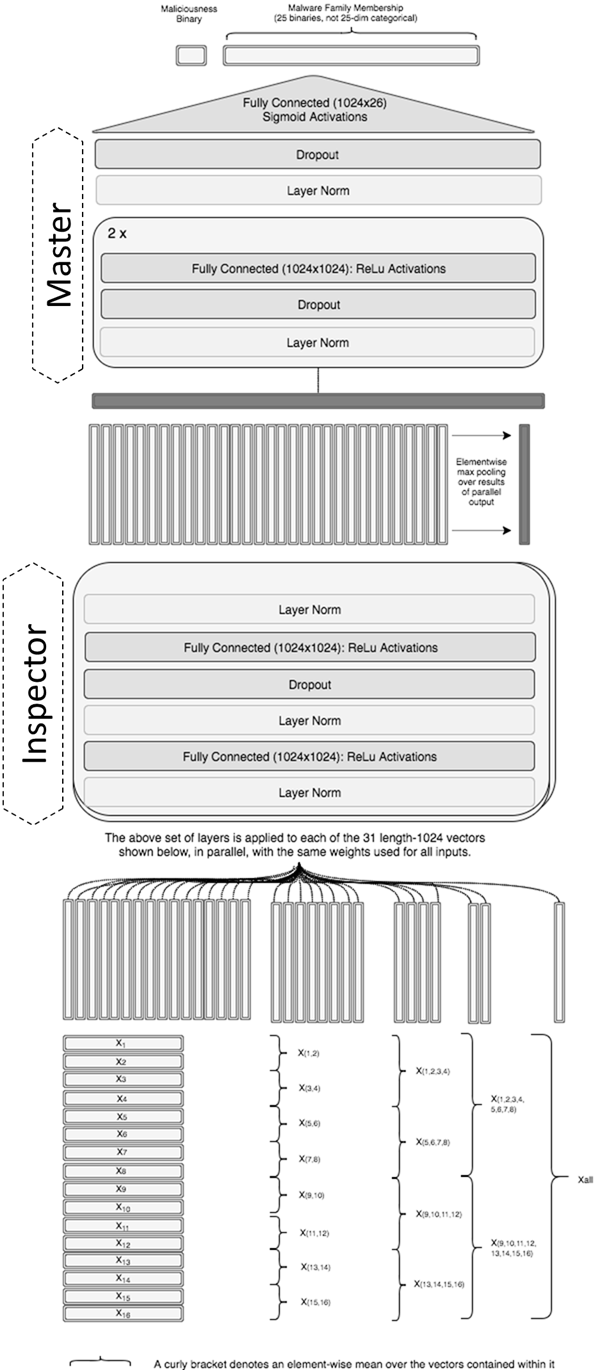}
      \caption{The architecture of our proposed hierarchical inspector approach}
      \label{fig:arch_figure}
\end{figure}

\subsection{Approach}
\label{sec:approach}
Our approach involves a feature extractor, which parses out a sequence of tokens from HTML documents, and a neural network model, which makes classification decisions by using a shared-weight examination of the features at hierarchical levels of aggregation.  The neural network includes two logical components:
\begin{itemize}
\item The first component is an \emph{inspector}, which applies weights at hierarchical spatial scales and aggregates information about the document into a 1024-length vector.  

\item The second component is a \emph{master network} which uses the output of the inspector network to make a final classification decision.  
\end{itemize}
The inspector and master components of the network are optimized jointly through back-propagation.  Below we walk through how each of these logical components contribute to our overall system.
%
%
%
%
%
%
%
%
%
%
%
%
%
%
%
%
%
%

\subsubsection*{Feature extraction}
\label{sec:features}
Our system's workflow begins by extracting a sequence of character-string tokens from HTML files.  First we tokenize the target document with the following regular expression: \verb!([^\x00-\x7F]+|\w+)!, which splits the document along non-alphanumeric word boundaries.  Then we divide the token stream up into 16 sequential chunks of equal length - where length is defined as number of tokens - including fewer tokens in the last chunk if the number of tokens in the documents is not divisible by 16. A reference implementation of the tokenization and chunking function is given as Python code in the Appendix of this paper under the heading \textit{TokenizeChunk}.

Next we use a modified version of the hashing trick, with 1024 bins, to create a bag-of-words style representation for each chunk.  We use a technique introduced in \cite{saxe2015deep} to modify bin placement so that it's a function both of feature hash and token length (a Python reference implementation is given as \textit{TokenLengthHash} in the Appendix). The result of this workflow, in which we tokenize, and then divide into 16 equal length token chunks, and then feature hash each token chunk into 1024 bins, is a 16x1024 tensor representing a sequence of bags-of-tokens extracted from an HTML document, where each element in the sequence represents an aggregation over a contiguous $\frac{1}{16}$ of the input document.
%
%
%
%
%
%
%
%
%
%

\subsubsection*{Inspector}
%
%
\label{sec:inspector}
Once we have a feature representation for an HTML document, we input that representation into our neural network, which we depict in Figure \ref{fig:arch_figure}.  As shown in the Figure, the first step in our computational flow is to create a hierarchical representation of our sequence of bags of tokens, in which we collapse our initial 16 bags of tokens into 8 bags of tokens, collapse these 8 bags of tokens into 4, those 4 into 2, and those 2 into 1, such that we obtain multiple bag-of-tokens representations capturing token occurrences at multiple spatial scales.  This collapsing process works by averaging windows of length 2 and step size 2 over the original 16 bags of tokens, and then doing this recursively until we get to a single bag of tokens. An important distinction to note here is that, by averaging, rather than summing, the token counts, we keep the norm of each representational level the same within a given document (though it can vary between documents of different overall length).
%
%
%
%
%
%
%
%
%
%

Once the inspector has created this hierarchical representation, it proceeds to visit each node in the tree of aggregations and compute an output vector.  As shown in Figure \ref{fig:arch_figure}, the inspector is a feed forward neural network with two fully connected layers which each have 1024 ReLU units.  We use layer normalization \cite{ba2016layer} to guard against vanishing gradients and dropout \cite{srivastava2014dropout} to regularize the inspector.  We use a dropout rate of 0.2.
%
%
%
%
%
%
%
%

To compute a 1024-dimensional vector output from the inspector after it has visited each node, we take the maximum activation from each of its 1024 output neurons across the 31 outputs produced from the 31 aggregated chunks. This causes the final vector representation of the document to be the maximum output of each neuron in the final layer of the inspector, given all of its activations over all nodes in the hierarchy.  Intuitively, this should promote the output vector capturing patterns that most closely match known template features useful in predicting malicious content, regardless of where they appear in the document, or how long the overall document is.
%
%
%

\subsubsection*{Master}
%
%
\label{sec:master}

Once the inspector has computed its 1024-dimensional output vector over the target document, this vector is input into the “master” component of our model.  As shown in Figure \ref{fig:arch_figure}, the master is implemented as a feed forward neural network with two logical fully-connected blocks, where each fully connected layer is preceded by layer normalization and dropout.  As in the case of the inspector, here we use a dropout rate of 0.2.
%
%
%
%
%
%

The final layer of the model consists of 26 sigmoid units corresponding to 26 detection decisions we make about documents.  One of these sigmoids is devoted to determining if the target document is malicious or benign.  The other 25 sigmoids detect a variety of informative tags, such as whether the document is a phishing document or an instance of an exploit kit.  To train the model, we use binary cross-entropy loss on each of these sigmoid outputs and then average the resulting gradients to compute parameter updates. In this paper, we emphasize evaluating the accuracy of the good vs. bad sigmoid output, but touch on our performance on these other outputs below as well.
%
%
%
%
%
%

\begin{figure}
%
%
%
      \includegraphics[width=\linewidth]{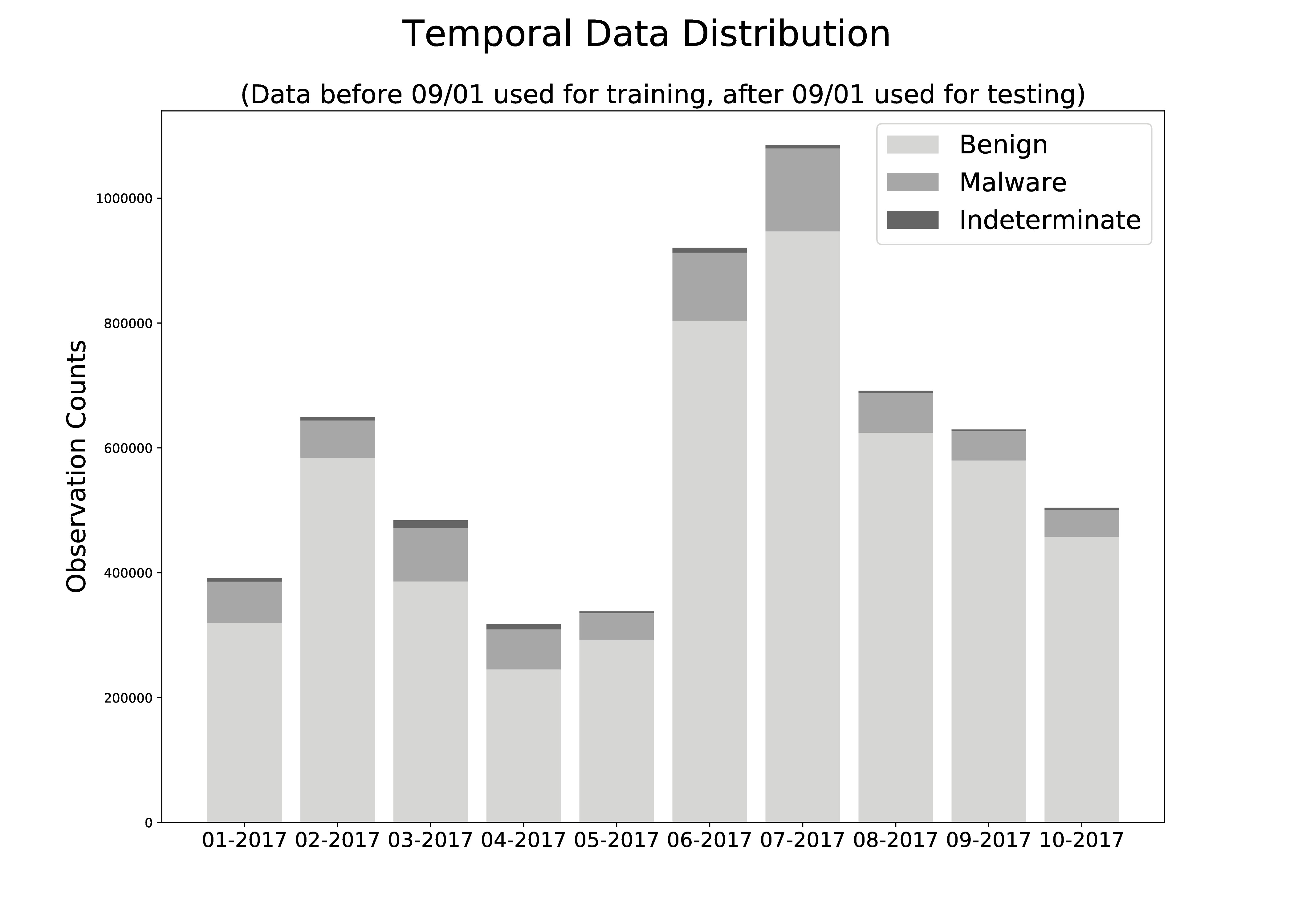}
      \caption{A time histogram showing when the samples in our training and validation sets were first seen on VirusTotal}
      \label{fig:data_dist_plot}
\end{figure}

\section{Evaluation}
\label{sec:evaluation}
We tested our approach in two ways.  First, we compared it to a number of bag-of-words style baselines which represent a standard document classification approaches.  Second, we modified the architecture in a variety of ways to test whether our model design choices were contributing to improved accuracy.  We did not directly compare our approach to approaches that involve complex parsing of or emulation of web content, because the performance overhead of these approaches placee them out of scope of our research goals, which were to create a fast web content detection model that can operate on web content observed on firewalls and endpoints.
%
%
%
%

Below we describe our experimental dataset, our baselines, and our model modifications and then introduce and discuss our results.

\subsubsection*{Experimental dataset and evaluation approach
}
We tested our approach on data collected from the threat intelligence site VirusTotal.  VirusTotal receives tens of thousands of new HTML files per day, scanning them with 60 web threat scanners from dozens of security vendors.  Sophos subscribes to VirusTotal's paid threat intelligence service, and as part of this subscription we receive every HTML file submitted to VirusTotal with its corresponding scanner results.  The experimental dataset used in this paper was collected from the VirusTotal feed in the first 10 months of 2017, as shown in Figure \ref{fig:data_dist_plot}.  

Files were uniquely identified based on SHA256, and our training/testing splits were computed on the basis of the time that the file was first reported on VirusTotal.  This process ensures that a) our training and test sets are distinct (as a later submission of an identical HTML file would be interpreted as a resubmission of that file, and we would ignore it in favor of the earlier version) and b) that our training and testing procedure at least approximates a real-world deployment scenario, as discussed in more detail below.   
%
%
%
%

Our labeling strategy is to use the scanner ensemble's response to a given file as the input to a labeling rule, which derives binary good versus bad labels from these data.  Specifically, our labeling rule defines benign HTML files as files receiving 0 detections from security vendors, and malicious files as files receiving 3 or more detections.
%
%

Additionally, we discard files that received 1 or 2 detections and did not use them in our research, because we consider this small number of detections to mean that the security industry is still uncertain as to whether they are malicious or benign. As can be seen in Figure \ref{fig:data_dist_plot}, these indeterminate files represent a very small proportion of the overall files from this time period. 
%
%
%
%

This labeling approach carries the risk that our approach will simply memorize the knowledge of security vendor products, as opposed to learning truly novel detection capabilities that detect malware the vendor community would have missed.  In this work, we test this using a historical simulation procedure, defined as follows:

\begin{enumerate}
\item We train our models on web content files first seen in the VirusTotal feed before some time \textit{t}.
\item We evaluate our models on web content files first seen in the two months after time \textit{t}, using the latest vendor labels.
%
%
\end{enumerate}

This evaluation procedure mitigates the problem of assessing our ability to detect malicious web content the vendor community would have missed.  This is because we test on files we haven't seen in training, but which the security vendor community has time to “catch up to” through detection rule and blacklist updates, thereby simulating the problem of detecting 0-day malicious content.
%
%
%
%
%
%

In other words, insofar as a machine learning model is able to correctly predict the label of future web content the vendor community has had time to blacklist or write detection rules for, we believe there is at least circumstantial evidence that our approach also has the ability to detect malicious web content the vendor community can be expected to miss entirely.

To provide more direct evidence of the efficacy of our proposed approach, we also hand-inspected samples that our labeling strategy marked as benign but for which our proposed model assigned a high probability of maliciousness, finding that a majority of these supposed false positives were either clearly malicious or junk content.  We discuss these results below.

\section{Experiments}
Aside from evaluating our proposed model, we conducted five additional experiments, three of which test the efficacy of our approach against alternative models, and two of which explore the inner workings of our model.  To test the efficacy of our model architecture we conducted the following three experiments. For all neural network models, models were trained using the Adam optimization scheme, balanced batches of size 64, and early stopping based on validation set performance.  We enumerate our test approaches below:
%
%
%
%

\subsubsection*{LR-BoT}
Elastic Net regularized logistic regression on bag of token features.  Here we used the tokens we extract for our full model, but feature hash them into a 16284-length vector.  We chose to use a 16284 input in this and other baseline experiments since it is the same dimensionality as the 16x1024 representation we use in our proposed model, and thus approximates an apples-to-apples comparison.  The LR-BoT test model provides a straightforward bag-of-tokens linear document classification approach as a point of comparison.  We determined the $\lambda_{1}$
%
%
%
%
%
 and $\lambda_{2}$ parameters for the L1 and L2 penalties via grid search.

\subsubsection*{FF-BoT}
A feed-forward architecture using the 16284-length feature hashed bag of tokens feature representation used above.  This test model provides a straightforward deep learning bag-of-words baseline as a point of comparison with our proposed approach.
%
%

\subsubsection*{XGBoost-BoT}
A gradient boosted decision tree (XGBoost) model using the same feature input as FF-BoT.

Separate from our baseline experiments, we performed several architecture modifications to determine the contribution of the design decisions we made in creating our proposed model, which we list here:

\subsubsection*{FlatSequential}
A variant of our proposed model with the average pooling step removed, such that the inspector only sees the leaf nodes of the tree. Put another way, out of the 31 aggregated representations input to the champion model, this model only sees the 16 sequential chunks, and none of the larger aggregated windows. This model tests the gain in performance provided by examining HTML documents at hierarchical spatial scales.

\subsubsection*{FlattenedFF}
A simple feed forward neural network that uses the same feature representation as our selected model.  However, rather than applying the shared-weight inspector at each step, we simply rasterize the 16x1024 sequential bag of tokens vectors into a single, 16284-length vector and input that into a feed forward neural network. This experiment evaluates the performance gain of using a shared-weight inspector at all, relative to a dense first layer.

\section{Results}
\label{sec:results}
%
%
\begin{figure}
      \includegraphics[width=\linewidth]{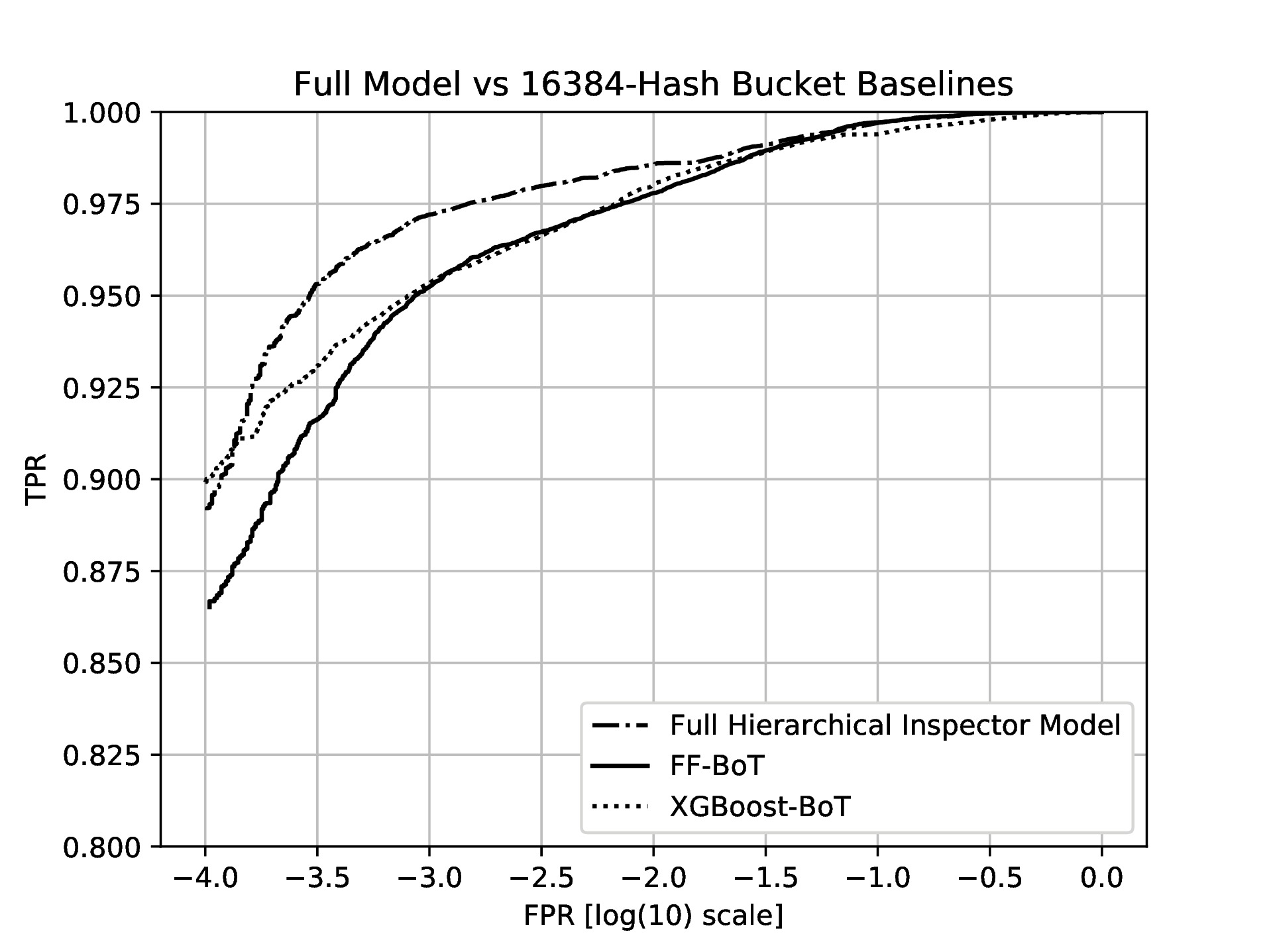}
      \caption{ROC curves showing the performance of our model versus two bag-of-tokens style baselines}

      \label{fig:model_v_baselines}
\end{figure}

\begin{figure}
      \includegraphics[width=\linewidth]{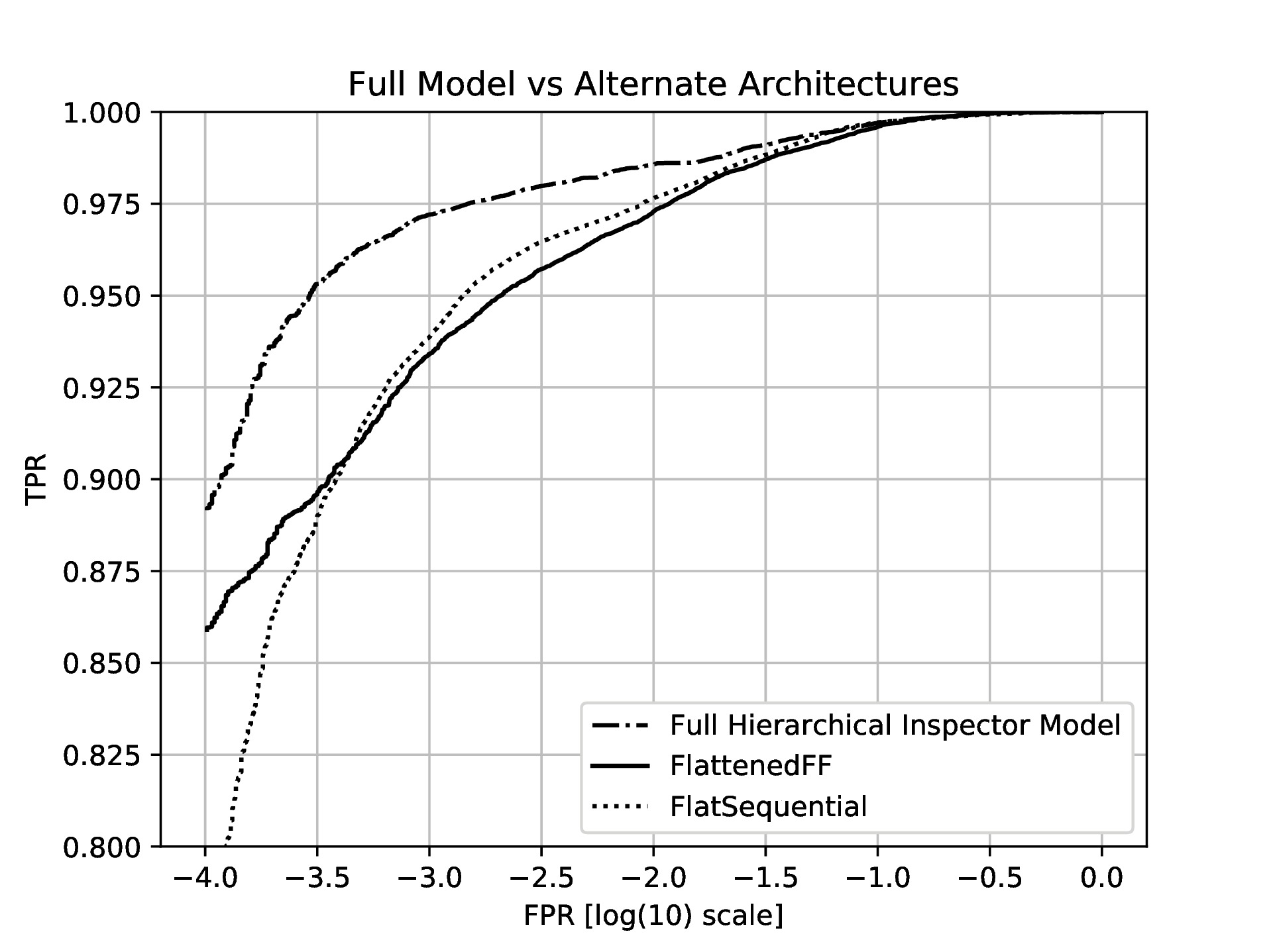}
      \caption{ROC curves showing the performance of our model versus two modified models}
      \label{fig:model_v_alt_arch}
\end{figure}

\begin{table}[]
\centering
\caption{Detection rates for different malware families, as well as the percentage of malware samples in which that tag appears}
\begin{tabular}{llr}
\textbf{Family Category}   & \textbf{DR@10e-3} & \textbf{Prevalence (\%)} \\
Code Injection XSS         & 0.999             & 16.1              \\
Browser Exploit            & 0.998             & 14.4             \\
iFrame Mischief            & 0.998             & 14.9             \\
Malicious Browser Redirect & 0.997             & 3.3               \\
Blackhat SEO               & 0.995             & 49.6              \\
Ramnit Malware Family      & 0.995             & 39.7              \\
Fake JQuery                & 0.977             & .5               \\
\textbf{All Malware}       & \textbf{0.972}    & \textbf{100}                   \\
Facebook Hacking           & 0.971             & 13.8             \\
Changes Browser Startpage  & 0.937             & 5.2               \\
Ransomware                 & 0.931             & 1.2               \\
Auto Click                 & 0.902             & .6                \\
Phishing                   & 0.895             & .5             
\end{tabular}
\label{tab:categoryperformance}
\end{table}

Figures \ref{fig:model_v_baselines.jpg} and \ref{fig:model_v_alt_arch} give our experimental results as ROC (receiver operating characteristic) curves, which show the trade-off between true positive rate (y-axis) and false positive rate (x-axis) as we adjust our detection threshold.  Figure \ref{fig:model_v_baselines} compares our baselines, FF-BoT and XGBoost-BoT, with our proposed approach, and Figure \ref{fig:model_v_alt_arch} compares our architecture modifications with our proposed approach.  Excluded from these figures is our linear baseline experiment, LR-BoT, which achieved a dramatically worse result than the rest of our experiments (i.e. 10\% detection rate at a 0.1\% false positive rate).

Inspecting the ROC curves in Figure \ref{fig:model_v_baselines}, we see that our proposed approach outperforms our baseline models.  If we compare these models' relative performance at a fixed false positive rate of 0.1\%, we see that our proposed approach, FF-BoT, and XGBoost-BoT achieve detection rates 97.2\%, 95.2\%, and 95.4\% respectively.  Posed in terms of false negative rates, our approach achieves a false negative rate of 2.8\% whereas our baselines' false negative rates are roughly double that, at 4.8\% and 4.6\%.  It is also worth noting that the overall ROC curve for our proposed model is significantly better than that of our baselines.

Interestingly, our FF-BoT baseline, which underperforms relative to our proposed model, has far more parameters (about 20 million) than our proposed model (about 4 million).  This suggests that our proposed approach captures a far more efficient representation of malicious HTML documents, thanks to the fact that our inspector uses the same parameters over every spatial context it examines in our hierarchical representation.

Inspecting the ROC curves in Figure \ref{fig:model_v_alt_arch}, we see that both the “hierarchical inspector” approach, in which a feed forward block is applied, with shared weights to each node in our bag of tokens hierarchy, outperforms variations on this architecture. FlattenedFF, which inspects our sequential bag-of-tokens representation with separate weights for each chunk achieves a 93.4\% detection rate, and FlatSequential, which is identical to our proposed model but without the average pooling step, achieves a 93.4\% detection rate as well.  Additionally the overall ROC curves for the test models are significantly worse than for our proposed approach.

The fact that our proposed approach beats FlatSequential is interesting, because it shows that inspecting content at multiple spatial scales is essential to achieving good accuracy.  Similarly, our approach beating FlattenedFF demonstrates that our inspector's usage of the same parameters upon inspection of every spatial context and scale is essential to yielding high detection accuracy, since FlattenedFF uses separate weights for every spatial context and achieves a worse result.

To get a better understanding of what our model learned, we took the malware family tags - earlier used as auxiliary targets - and subdivided true malware samples up according to which tags were attached to them. Based on a threshold corresponding to a 10e-3 global false positive rate, we calculated the overall detection rate - 97.2\% - and compared that to the detection rate for malware tagged as each of our most prevalent families. This comparison, shown in Table \ref{tab:categoryperformance}, demonstrates that our final model had the highest success with code injection, browser exploit, and iFrame manipulation attacks, and had the most trouble with phishing websites. A note about the percentages in the prevalance table: these tags are not mutually exclusive, which is why the prevalance rates do not add up to 100. 

In addition to large scale validation, we also checked to see whether or not cases where our model disagreed with vendor-based labels proved that our model had actually detected unknown malicious web content that the security community missed.  To do this analysis we inspected the top 20 highest scoring test examples from our validation set that the vendor community unanimously marked as benign, and found that 13 of the 20 were in fact malicious or potentially unwanted, and 7 of them were false positives.  Of the examples that we found were malicious, 3 were the alarm pages from web content blockers, not malicious themselves but which indicate maliciousness, 3 were malicious fake JQuery libraries, 1 was Javascript that drops a fake svchost.exe file on disk, 1 was a Facebook “clickjacking” page, 1 contained code to perform a drive-by-download, and 2 were Viagra spam.  This analysis verifies that our model is capable of generalizing beyond label noise and identifying previously unidentified malicious content.

\section{Conclusion}
\label{sec:Conclusion}
One consistent theme in the recent successes of applied deep learning has been the value of using known structural features of a domain - such as locality-based features and translational invariance in the case of images -  to channel and constrain a model's learning ability in the direction of that known structure. While this paper's domain is more specialized, it continues in that same tradition of using domain knowledge to provide the useful inductive bias of hierarchical spatial scaling, which we believe makes our model more effective at handling detection problems within documents of potentially widely-varying size. In this paper, we achieve strong performance of 97.5\% detection at a 0.1\% false positive, and even identify malicious content not previously caught by the vendor community, with a purely token-based static approach that avoids the need for complex parsing or emulation systems. This result gives us greater confidence that deep learning systems can learn high quality internal representations of fairly raw web content inputs that outperform hand-crafted features, and, more broadly, that deep learning approaches have a promising future in the detection of malicious web content.

\section*{Acknowledgment}

We would like to thank the other members of our research team, including Alex Long, Ethan Rudd, Felipe Ducau, Konstantin Berlin, Madeleine Schiappa, and Omar Alrawi, for support and feedback during the research and writing that went into this paper.

\bibliography{main}
\bibliographystyle{ieeetr}
\section*{Appendix}

Here we provide short Python reference implementations of the functions that produce the input to our proposed architecture, including \textit{TokenizeChunk}, which tokenizes a target web document and splits it into chunks, and \textit{TokenLengthHash}, which creates feature vectors using a modified version of the hashing trick for each of these chunks.

\subsection*{TokenizeLengthHash}
\begin{verbatim}

import re
import murmur # the murmur hashing library

def TokenizeLengthHash(data, steps=16, dims=16384):
    feats = re.findall(r"([^\x00-\x7F]+|\w+)", data)
    final_feats = []
    for feat in feats:
        loglength = int(min(8, max(1, math.log(len(feat), 1.4)))) - 1  # 0-7
        shash = murmur.string_hash(feat) % (dims / steps / 8)
        final_feats.append(loglength * (dims / steps / 8) + shash)
    return final_feats
    
\end{verbatim}

\subsection*{TokenizeChunk}
\begin{verbatim}

import numpy as np

def TokenizeChunk(data, steps=16, dims=16384):
    data = TokenizeLengthHash(data, steps=steps, dims=dims)
    ret = []
    stepsize = int(len(data) / float(steps))

    for percent in np.arange(0, 1, 1 / float(steps)):
        idx = int(len(data) * percent)
        unq, cnt = np.unique(data[idx:idx + stepsize], return_counts=True)
        newarray = np.zeros(dims / steps)
        for v, c in zip(unq, cnt):
            newarray[v] = c
        ret.append(newarray)
    return ret
    
\end{verbatim}

\end{document}